\documentclass[11pt]{article}
\usepackage{graphicx} 
\usepackage{amsmath}
\usepackage{lipsum}
\usepackage{pdfcomment}

\usepackage{tikz}
\usetikzlibrary{arrows.meta, positioning, shapes.geometric, fit}
\usepackage{caption}
\usepackage{amsmath}
\usetikzlibrary{positioning}
\usetikzlibrary{arrows.meta, positioning}

\usepackage[table,xcdraw]{xcolor} 
\usepackage{subcaption}
\usepackage[utf8]{inputenc}
\usepackage{longtable}
\usepackage{graphicx}
\usepackage{float}
\usepackage{booktabs}
\usepackage{float} 
\usepackage{geometry}
\usepackage{lineno}
\usepackage{authblk}
\usepackage{tcolorbox}
\usepackage{rotating}
\usepackage{booktabs}
\usepackage{threeparttable}
\usepackage{geometry} 
\usepackage{lscape}
\usepackage{siunitx}  
\usepackage{adjustbox} 
\usepackage{pgfplots}
\usepackage{caption}
\usepgfplotslibrary{groupplots}

\usepackage[numbers,sort&compress]{natbib}

\usepackage{hyperref}
\hypersetup{
  colorlinks=true,
  linkcolor=blue,       
  citecolor=blue,       
  urlcolor=blue         
}

\usepackage{cite}

\usepackage{array}
\usepackage{multirow}

\usepackage{amssymb}
\usepackage{pifont}

\usepackage[colorinlistoftodos]{todonotes}

\usepackage{wrapfig}

\usepackage{cite}

\usepackage{amsthm}

\usepackage{titlesec}
\usepackage{titletoc}
\titleclass{\subsubsubsection}{straight}[\subsection]
\newcounter{subsubsubsection}[subsubsection]
\renewcommand\thesubsubsubsection{\thesubsubsection.\arabic{subsubsubsection}}
\titleformat{\subsubsubsection}{\normalfont\normalsize\bfseries}{\thesubsubsubsection}{1em}{}
\titlespacing*{\subsubsubsection}{0pt}{3.25ex plus 1ex minus .2ex}{1.5ex plus .2ex}
\titleformat*{\section}{\fontsize{11}{12}\selectfont\bfseries}
\titleformat*{\subsection}{\fontsize{11}{12}\selectfont\bfseries}
\titleformat*{\subsubsection}{\fontsize{11}{12}\selectfont\bfseries}
\titleformat*{\paragraph}{\fontsize{11}{12}\selectfont\bfseries}
\titleformat*{\subparagraph}{\fontsize{11}{12}\selectfont\bfseries}

\titlecontents{subsubsubsection}
[8em] 
{\small}
{\hyperlink{toc.\thecontentslabel}{\thecontentslabel.} }
{}
{\ \titlerule*[.5pc]{.}\contentspage}

\titlespacing*{\subsubsubsection}{0pt}{3.25ex plus 1ex minus .2ex}{1.5ex plus .2ex}

\usepackage{caption}
\usepackage{microtype}

\usepackage{enumitem}
\setlist{itemsep=0em}

\usepackage{fancyhdr} 
\usepackage{lastpage} 

\usepackage{listings}
\usepackage{xcolor}

\definecolor{codegreen}{rgb}{0,0.6,0}
\definecolor{codegray}{rgb}{0.5,0.5,0.5}
\definecolor{codepurple}{rgb}{0.58,0,0.82}
\definecolor{backcolour}{rgb}{0.95,0.95,0.92}

\lstdefinestyle{mystyle}{
    backgroundcolor=\color{backcolour},   
    commentstyle=\color{codegreen},
    keywordstyle=\color{magenta},
    numberstyle=\tiny\color{codegray},
    stringstyle=\color{codepurple},
    basicstyle=\ttfamily\footnotesize,
    breakatwhitespace=false,         
    breaklines=true,                 
    captionpos=b,                    
    keepspaces=true,                 
    numbers=left,                    
    numbersep=5pt,                  
    showspaces=false,                
    showstringspaces=false,
    showtabs=false,                  
    tabsize=2
}

\title{A Bilevel Optimization Framework for Adversarial Control of Gas Pipeline Operations}

\author[1]{Tejaswini Sanjay Katale}
\author[2]{Lu Gao}
\author[3]{Yunpeng Zhang}
\author[2]{Alaa Senouci}

\affil[1]{Department of Computer Science, University of Houston}
\affil[2]{Department of Civil and Environmental Engineering, University of Houston}
\affil[3]{Department of Information Science Technology, University of Houston}

\date{}

\begin{document}

\maketitle

\abstract{
Cyberattacks on pipeline operational technology systems pose growing risks to energy infrastructure. This study develops a physics-informed simulation and optimization framework for analyzing cyber–physical threats in petroleum pipeline networks. The model integrates networked hydraulic dynamics, SCADA-based state estimation, model predictive control (MPC), and a bi-level formulation for stealthy false-data injection (FDI) attacks. Pipeline flow and pressure dynamics are modeled on a directed graph using nodal pressure evolution and edge-based Weymouth-type relations, including control-aware equipment such as valves and compressors. An extended Kalman filter estimates the full network state from partial SCADA telemetry. The controller computes pressure-safe control inputs via MPC under actuator constraints and forecasted demands. Adversarial manipulation is formalized as a bi-level optimization problem where an attacker perturbs sensor data to degrade throughput while remaining undetected by bad-data detectors. This attack-control interaction is solved via Karush–Kuhn–Tucker (KKT) reformulation, which results in a tractable mixed-integer quadratic program. Test gas pipeline case studies demonstrate the covert reduction of service delivery under attack. Results show that undetectable attacks can cause sustained throughput loss with minimal instantaneous deviation. This reveals the need for integrated detection and control strategies in cyber-physical infrastructure.
}


\section{Introduction}\label{introduction}

Critical pipeline infrastructure networks are the backbone of modern energy transportation, which enables the large-scale delivery of oil, gas, and refined petroleum products over vast geographic regions. These networks, composed of interconnected pipelines, pump stations, valves, and storage facilities, operate continuously to meet dynamic energy demands. Their reliable performance is essential for economic stability, national security, and the functioning of industrial and consumer sectors \citep{Chen2021PipelineSafetySecurity}.

Over the past two decades, the digitalization of pipeline operations through Supervisory Control and Data Acquisition (SCADA) systems, Industrial Control Systems (ICS), and distributed IoT-based sensors has enhanced operational efficiency, improved situational awareness, and enabled predictive maintenance \citep{Enemosah2024SCADAIoT}. However, this integration of cyber and physical components has also expanded the potential attack surface, which exposes critical pipeline systems to sophisticated cyber-physical threats. Malicious actors can exploit vulnerabilities in both information technology (IT) and operational technology (OT) domains, which has the potential to cause severe disruptions to energy supply chains \citep{Kayan2021ICPSecurity}. 

Real-world incidents have underscored the severity of such risks. For example, the 2021 Colonial Pipeline ransomware attack demonstrated that compromising IT assets, even without directly tampering with OT controls, can lead to precautionary shutdowns of physical operations. This resulted in fuel shortages, price spikes, and cascading supply chain effects \citep{tsvetanov2021effect}.  Similarly, targeted manipulation of OT components, such as pumps and valves, can disrupt hydraulic stability, reduce throughput, and damage physical assets. These highlight the urgent need for analytical and simulation tools to assess pipeline system resilience under cyber-attack scenarios. 

While prior studies have explored cyber-physical vulnerabilities in industrial systems, research specifically addressing pipeline infrastructure networks remains relatively limited. Existing approaches often focus exclusively on either cyber-attack detection or physical flow modeling, without integrating both aspects into a unified framework. As a result, there is a lack of simulation platforms capable of representing realistic hydraulic dynamics alongside diverse cyber-attack vectors. This gap limits the ability of operators, policymakers, and security analysts to anticipate attack impacts, design robust countermeasures, and evaluate recovery strategies. In this study, we propose a physics-informed, graph-based framework for evaluating cyber-attack impacts on pipeline infrastructure networks. The framework models pipeline hydraulics coupled with discrete-time network flow dynamics. A case study on a test pipeline network illustrates how disruptions propagate through the network.

\section{Literature Review}

\subsection{Cybersecurity in Critical Infrastructure Systems}

Advances in sensing, communication, and automation have transformed traditional infrastructure systems into highly interconnected, intelligent networks. For example, across diverse sectors such as transportation, energy, healthcare, and the built environment, infrastructure systems are adopting advanced technologies including connected and autonomous vehicles, real-time monitoring and control, Internet of Things (IoT) devices, and digital modeling to enhance operational intelligence and connectivity \citep{song2017smart,madireddy2025large,gao2025exploring,lebaku2025cybersecurity,silwal2024assessing,chenchu2025signals}. These smart and connected infrastructures promise significant gains in efficiency and safety. However, they also introduce complex cyber-physical vulnerabilities \citep{CISA2023}. Malicious actors can exploit weaknesses in IoT devices, communication protocols, and autonomous control systems to disrupt services, cause physical damage, or compromise safety. Beyond detection and control methods, practical deployment should align with security–privacy frameworks and interoperable industrial AI platforms \citep{alonso2024interoperable}. Recent incidents illustrate these risks, including the 2016 ransomware attack on the San Francisco Municipal Transportation Agency that disrupted fare collection and transit operations \citep{BBC2016SFMTA}, the 2021 Colonial Pipeline ransomware attack that halted fuel delivery across much of the U.S. East Coast \citep{CISA2021ColonialPipeline}, and the 2020 ransomware incident at Vermont Medical Center that delayed surgeries and disabled electronic medical records \citep{HHS2020Vermont}. Most recently, in July 2025, a coordinated attack struck the City of St. Paul, Minnesota’s municipal information systems, forcing officials to shut down critical IT infrastructure \citep{Reuters2025StPaul}. These incidents demonstrate how highly interconnected infrastructures create intricate cyber-physical dependencies, where a digital breach can cascade into operational paralysis and pose significant public safety risks.

Previous studies have identified various cyberattack methods in OT and ICS \citep{Sridhar2012ProcIEEE}. One widely studied type of cyberattack is reconnaissance and lateral movement, in which attackers begin by scanning and analyzing the network to gather information about its structure, devices, and software. After gaining initial access, they move from one part of the system to another by exploiting outdated technologies and the lack of proper separation between enterprise and control networks, aiming to reach critical components without being detected \citep{Knowles2015IJCIP}. False-data injection is a commonly studied attack technique in which adversaries modify sensor measurements to mislead the system’s state estimation, causing the controller to make incorrect decisions while passing standard error checks \citep{Kosut2011TSG}.  
Replay attacks involve recording legitimate sensor or control signals and then resending them at a later time, which allows attackers to perform unauthorized actions while the system continues to observe data that appears valid \citep{Teixeira2015Automatica}.  
Command and logic manipulation refers to altering control instructions, setpoints, or the internal logic of programmable devices, as demonstrated by malware that rewrites industrial controller code to trigger physical damage without immediate detection \citep{Langner2011}.  
Denial-of-service and resource-exhaustion attacks reduce system availability by overwhelming communication channels, computation units, or control loops, which disrupts real-time feedback and prevents operators from monitoring or intervening effectively \citep{Dibaji2019ARC}.  
Stealthy attacks remain active in the system without triggering alarms by introducing subtle changes that preserve normal operating patterns, making it difficult to detect them using conventional monitoring methods \citep{Pasqualetti2013TAC}.  


\subsection{Pipeline Network Modeling and Control}

Pipeline transmission systems are typically represented as graphs whose edges denote pipes and whose nodes denote junctions, supplies, withdrawals, compressors, and regulators, with nodal coupling conditions enforcing mass conservation and element-specific pressure relations \citep{Osiadacz1984IJNMF}. Pipeline networks are commonly modeled by applying physical conservation laws to describe the dynamic relationships among pressure, flow, and gas density. On each pipe, gas transport is typically formulated using one-dimensional compressible flow equations that include the continuity equation for mass conservation and a momentum equation that captures pressure gradients, inertia, and friction effects \citep{Thorley1987IJHFF}. The Darcy–Weisbach equation is frequently used to quantify pressure loss due to friction, expressed as a function of velocity, pipe roughness, and diameter \citep{Osiadacz1987GPE}. These fundamental equations relate the temporal and spatial variation of pressure and flow rate along each pipeline segment. In cases where temperature variations significantly affect gas behavior, an additional energy balance equation is introduced to model thermal dynamics and heat exchange with surrounding soil \citep{Chaczykowski2010APM}.

In pipeline networks, Kalman filter-based approaches are widely employed to estimate the distributed hydraulic state by integrating sparse sensor measurements with physical models. These methods rely on variants of the Kalman filter to assimilate telemetry data and infer unmeasured pressures and flows while accounting for noise and model uncertainty \citep{Durgut2016JNGSE}. For example, extended Kalman filters (EKF) are commonly used to handle the nonlinearities in the pipe dynamics by linearizing the system around current estimates \citep{Liu2025EKFPipeline}. When high-fidelity modeling is required, unscented Kalman filters (UKF) offer improved performance by capturing nonlinear transformations without explicit linearization \citep{julier2004unscented}. These estimation frameworks can also incorporate composition-dependent variables by augmenting the state vector with gas species balances, enabling joint inference of hydraulic and chemical parameters \citep{Chaczykowski2018JNGSE}. In operational settings, residuals between predicted and observed values are often monitored to detect anomalies such as leaks or faults, further demonstrating the utility of Kalman filtering as both a state estimator and a diagnostic tool \citep{Bar-Shalom2001Estimation}. 

Model predictive control (MPC) has been widely applied to optimize gas pipeline operations by adjusting compressor and valve actions over a receding horizon, while satisfying transient hydraulic constraints on pressures, flows, and actuators \citep{Bu2024CCE}. Variants such as tracking MPC and economic nonlinear MPC have been developed to update unmeasured states in real time and reduce energy and fuel costs, respectively, while recent work incorporates data-driven models to address plant–model mismatch and improve control under fully transient conditions \citep{Zhang2023CCE, Ghilardi2025CCE, Moetamedzadeh2019TIMC}. These control strategies rely on supervisory control and data acquisition (SCADA) systems, which collect real-time measurements and issue operational commands through networks of field sensors, remote terminal units, and centralized control centers \citep{Yadav2021IJCIP}. SCADA data supports state estimation using Kalman filter variants to infer pressures and flows at uninstrumented locations, feeding critical feedback signals into MPC \citep{Durgut2016JNGSE}. Additionally, SCADA historians and alarm systems enable leak detection by comparing real-time measurements with transient model predictions \citep{Lu2017Energies}, and machine learning methods have been applied to SCADA telemetry to detect rare cyber or process anomalies under class imbalance \citep{Choubineh2020IJCIP}. As SCADA adopts open protocols and IP networking, the expanded connectivity introduces new cybersecurity risks, making it vital to combine telemetry with physics-based models and residual analysis to enhance anomaly detection and reduce false alarms \citep{Alanazi2023COSE, Adegboye2019Sensors}.

\subsection{Cyber-Physical Modeling of Pipeline Attacks}

Prior work has modeled cyberattacks against pipeline SCADA telemetry using various mathematical and machine learning frameworks. For example, \citet{Choubineh2020IJCIP} introduced a cost-sensitive SCADA attack classifier that leverages Fisher’s discriminant analysis to correct extreme class imbalance on a virtual gas pipeline dataset. The modeling encodes misclassification asymmetry through class-dependent costs and forms linear discriminants on windowed telemetry vectors to separate benign and malicious events. \citet{Zheng2022Energy} proposed a deep anomaly detector for multi-product pipelines that exploits coupled spatial and temporal correlations in operations. The model constructs feature tensors over pipeline segments and time lags and trains a supervised network to capture coordinated deviations across stations. \citet{Xu2022ApplSci} designed a transformer-based generative adversarial network for SCADA time series that learns normal behavior and flags attacks via reconstruction discrepancies. The generator–discriminator pair uses attention to model long-range dependencies, and an anomaly score blends reconstruction error with discriminator confidence. \citet{Altaha2022Electronics} built a protocol-aware intrusion detector for DNP3 traffic by modeling function-code usage and sequencing patterns relevant to pipeline SCADA. The modeling derives statistical profiles over command types and inter-arrival timing and applies unsupervised clustering to expose protocol-level manipulations. 
\citet{Kim2023Sensors} presented a comparative benchmarking framework for ICS time-series detectors to guide model selection under operational variability. The framework standardizes preprocessing, sliding-windowing, and thresholding and reports metrics such as F1 and AUROC across representative operating regimes. \citet{Durgut2016JNGSE} applied a Kalman-filter-based state estimator to transient gas pipelines so that residuals between predictions and measurements act as physics-informed attack indicators. 

Another related line of work is secure state estimation under stealthy false data injection (FDI). The goal is to keep estimation errors bounded even when some sensors are arbitrarily compromised. Many approaches rely on attack sparsity and sensor reconfiguration. For example, adaptive switching observers can isolate corrupted channels when the number of attacked sensors remains below a detectability threshold \citep{AnYang_TAC2018}. Robust estimators with provable performance have also been designed by combining local observers, residual screening, and fusion to approach the fundamental limits under sparse sensor integrity attacks \citep{NakahiraMo_TAC2018}. On the adversary’s side, optimal linear deception strategies for remote state estimation have been analyzed to capture stealth constraints and the trade-off between attack impact and detectability \citep{GuoShiJohanssonShi_TCNS2017}. These studies complement our focus: instead of proposing a new secure estimator, we design stealthy attacks through a bilevel optimization program and measure their closed-loop impact (throughput and RMSE) under a standard EKF–MPC framework. Our setup can also serve as a benchmark environment for testing secure estimation methods under the same attack budget.

The modeling linearizes isothermal pipe dynamics around an operating point and calibrates process and sensor noise to reconstruct unmeasured pressures and flows. \citet{Isom2018CACE} combined an unscented Kalman filter with quadratic-program data reconciliation to fuse noisy measurements in gas pipeline networks. The model enforces nodal mass-balance and bound constraints while minimizing adjustment norms, yielding estimates robust to outliers and sensor faults. \citet{Marino2021CIE} proposed a cyber–physical resilience assessment that couples gas-transmission hydraulics with SCADA dependencies to quantify disruption and recovery. The modeling integrates network-flow or transient physics with a discrete-event layer for communication and control, producing service-loss and recovery-time metrics under cyber scenarios. \citet{Rezazadeh2019RESS} formulated a game-theoretic attacker–defender model for oil and gas pipeline security that allocates protective resources and evaluates adversarial incentives. The framework specifies payoff functions in terms of throughput loss and protection cost and computes equilibrium strategies over targets and countermeasures.  \citet{Fawzi2014TAC} constructed an optimization-based secure estimator that recovers system state under sparse adversarial sensor or actuator corruption. The model poses convex programs with sparsity-promoting penalties and provides identifiability conditions under which corrupted entries are isolated and states are consistently estimated. \citet{Teixeira2015Automatica} proposed a secure-control framework that formalizes replay, bias, and zero-dynamics attacks from resource-limited adversaries. The modeling characterizes reachable sets under constrained attack channels and derives detectability and performance bounds for feedback loops relevant to pipeline control. \citet{Pasqualetti2013TAC} contributed graph- and descriptor-system-based monitors for attack detection and identification in constrained networked dynamics. The approach uses structural left-invertibility and residual generators to localize compromised nodes and signals in differential–algebraic models akin to pipeline networks.

\subsection{Limitation of Existing Research and Motivation}

Despite substantial progress in cyberattack detection and modeling within pipeline SCADA systems, a key limitation of existing studies is the lack of a comprehensive modeling framework that connects the full process from sensor-level attacks to their downstream effects on estimation, control, and system performance. Many prior works focus on isolated components, such as anomaly detection in telemetry or analysis of specific attack types in static environments. However, they rarely simulate how malicious perturbations propagate through state estimation algorithms and influence real-time control actions and operational outcomes. This absence of an integrated dynamic framework prevents a full understanding of the operational consequences of cyber threats and limits the development of unified assessment and mitigation strategies.

To address this gap, the present study develops a closed-loop modeling and simulation framework that captures the complete impact of cyberattacks on pipeline network operations. By jointly modeling telemetry perturbations, Kalman-filter-based state estimation, and model predictive control under dynamic hydraulic constraints, the framework enables system-level evaluation of attack propagation and response. This unified approach facilitates vulnerability analysis, resilience testing, and control hardening for pipeline cyber–physical security.

\section{Methodology}

\begin{figure}[H]
    \centering
    \includegraphics[width=0.7\linewidth]{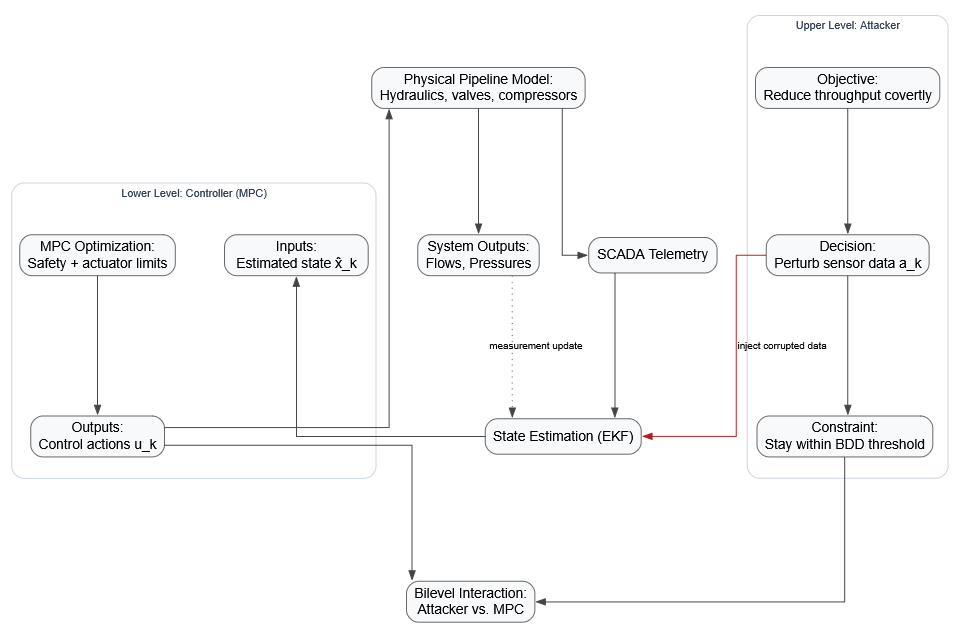}
    \caption{Overview of the Proposed Framework}
    \label{fig:overview}
\end{figure}

This section presents a dynamic modeling and simulation framework for petroleum pipeline networks under cyberattacks on operational technology systems (\autoref{fig:overview}). The framework captures the network topology, hydraulic and device relationships, supervisory control logic, and monitoring mechanisms, enabling the analysis of how malicious data injections or control manipulations propagate through the system and affect operations. The objective is to evaluate network vulnerability, quantify operational impacts, and assess the effectiveness of mitigation strategies.

The modeling is organized into three layers:  
(i) Network representation and hydraulics, a graph-based model of nodes and edges with associated flow and pressure relationships.  
(ii) Control and monitoring, a supervisory controller using Model Predictive Control (MPC) with state estimation from SCADA measurements.  
(iii) Optimization-based attack and control interaction, a bi-level formulation where the upper level (attacker) designs covert measurement perturbations to disrupt network performance, and the lower level (controller) responds optimally via MPC.

  
\subsection{Network topology and hydraulic modeling}

\subsubsection{Network representation}

Let $G = (\mathcal{V},\mathcal{E})$ denote a directed graph representing the pipeline network, where $|\mathcal{V}| = n$ is the number of nodes and $|\mathcal{E}| = m$ is the number of edges. The nodal pressure vector $\mathbf{p}(t) \in \mathbb{R}^n$ contains the pressures at each node at time $t$. The edge flow vector $\mathbf{q}(t) \in \mathbb{R}^m$ contains the mass (or standard volumetric) flow rates along the directed edges. The external injection vector $\mathbf{w}(t) \in \mathbb{R}^n$ specifies the supply or withdrawal of energy-carrying gas at each node, with positive values representing injection and negative values representing extraction. In this formulation, nodes correspond to junctions, sources (inlets), sinks (demands), or equipment interfaces, while edges correspond to physical pipelines or equipment connections. Pressures are defined at nodes, and flows are associated with edges.

The oriented incidence matrix $\mathbf{B} \in \mathbb{R}^{n \times m}$ encodes the network topology and the orientation of edges in the directed graph $G = (\mathcal{V}, \mathcal{E})$. Its entries are defined as
\begin{equation}
B_{i e} =
\begin{cases}
+1, & \text{if edge } e = (i \!\to\! j) \text{ is directed outward from node } i,\\[2pt]
-1, & \text{if edge } e = (j \!\to\! i) \text{ is directed inward to node } i,\\[2pt]
0,  & \text{if node } i \text{ is not incident to edge } e.
\end{cases}
\end{equation}
For an edge flow vector $\mathbf{q}(t) \in \mathbb{R}^m$, the product $\mathbf{B} \, \mathbf{q}(t)$ gives the net outflow at each node, with positive entries indicating net outflow and negative entries indicating net inflow.

For each node $i \in \mathcal{V}$, the equivalent nodal volume is defined as  
\begin{equation}
V_i = \frac{\pi}{8} \sum_{e \in \mathcal{N}(i)} D_e^{2} L_e,
\qquad
\mathbf{V} = \mathrm{diag}(V_1, \dots, V_n),
\end{equation}
where $\mathcal{N}(i)$ denotes the set of edges incident to node $i$, $D_e$ is the internal diameter of edge $e$, and $L_e$ is its length.  
This formulation assumes that each pipeline segment shares its physical volume equally between its two endpoint nodes, such that one-half of the volume $\frac{\pi D_e^{2}}{4} L_e$ is allocated to each node, giving $\frac{\pi}{8} D_e^{2} L_e$.  
The scalar $V_i$ represents the lumped line-pack capacity associated with node $i$, serving as a local storage proxy in the pressure–flow dynamics.  
The diagonal matrix $\mathbf{V}$ is subsequently used to scale nodal mass-balance equations into pressure-dynamics form.

\subsubsection{Edge flow models}
In pipeline network modeling, edge flow models describe the relationship between pressures at the endpoints of an edge and the resulting flow along that edge.
These models capture both passive flow in standard pipelines and active control behavior in equipment such as compressors and valves.

In the absence of active equipment such as compressors or control valves, the flow along a pipeline segment $e = (i \!\to\! j)$ is modeled using the quasi-steady isothermal compressible Weymouth-type relation \citep{menon2005gas}:
\begin{equation}
q_e(t) = \frac{1}{K_e} \, \mathrm{sgn}\!\big(p_i(t) - p_j(t)\big) \;
\sqrt{ \big| p_i^2(t) - p_j^2(t) \big| } .
\label{eq:weymouth}
\end{equation}
where,\\
$q_e(t)$ = the mass (or standard volumetric) flow rate along edge $e$;\\
$p_i(t)$, $p_j(t)$ = the pressures at the upstream and downstream nodes, respectively. \\
$\mathrm{sgn}(\cdot)$ = symbol ensures that flow is directed from higher to lower pressure;\\
$K_e$ = the composite hydraulic resistance, given by $K_e = \sqrt{\frac{16\,f_e\,c^2\,L_e}{\pi^2 D_e^{5}}}$, where $f_e$ is the Darcy-Weisbach friction factor, $D_e$ is the internal diameter of the pipe, $L_e$ is the pipe length, and $c$ is the isothermal speed of sound in the transported gas. 

For an equipment edge $e=(i\!\to\!j)$, a common example of a control-aware
constitutive relation (suitable for throttling devices such as control valves
or chokes) writes the squared-pressure drop with a control-dependent resistance:
\begin{equation}
q_e(t)
\;=\;
\frac{1}{\sqrt{\,w_e\big(\alpha_e(t);\theta_e\big)\,}}\;
\mathrm{sgn}\!\big(p_i(t)-p_j(t)\big)\;
\sqrt{\big|p_i^2(t)-p_j^2(t)\big|}.
\label{eq:equip-flow}
\end{equation}
Here $\alpha_e(t)\in[0,1]$ is the device control input (for example a valve
opening), $w_e(\alpha_e;\theta_e)>0$ is a resistance coefficient that decreases
monotonically with opening, and $\theta_e$ collects fixed device parameters
(such as valve $C_v$ curve, geometric limits, and calibrated loss factors).
Equation \eqref{eq:equip-flow} reduces to the standard Weymouth-type relation when $w_e(\alpha_e;\theta_e)$ is constant, and
they capture the expected behavior that smaller openings yield larger
resistance and lower flow for the same pressure drop.

\subsubsection{Nodal pressure dynamics}

Nodal pressure dynamics describe how the pressures at network nodes change over time in response to net inflows, withdrawals, and the storage capacity of connected pipelines. For each node, the net inflow from connected edges changes the amount of fluid stored locally in the surrounding pipes, which in turn changes the local pressure. This leads to the nodal pressure dynamics

\begin{equation}
\dot p_i(t)
\;=\;
c^2\,\frac{1}{V_i}\!\left(
\sum_{e\in\mathcal E_i^{\mathrm{in}}} q_e(t)
\;-\;
\sum_{e\in\mathcal E_i^{\mathrm{out}}} q_e(t)
\;+\; w_i(t)
\right),
\qquad i=1,\dots,n,
\label{eq:node-ode-scalar}
\end{equation}
where $p_i(t)$ denotes the nodal pressure at node $i$, $\dot p_i(t)$ denotes its time derivative. 
$\mathcal E_i^{\mathrm{in}}$ and $\mathcal E_i^{\mathrm{out}}$ are the sets of edges directed into and out of node $i$,
$q_e(t)$ is the flow on edge $e$ (positive in the edge's own direction),
$w_i(t)$ is the external injection ($>0$) or withdrawal ($<0$) at node $i$.
$V_i>0$ is the equivalent nodal volume,
and $c$ is the isothermal speed of sound.
The difference $\sum_{e\in\mathcal E_i^{\mathrm{in}}} q_e(t)-\sum_{e\in\mathcal E_i^{\mathrm{out}}} q_e(t)$ equals the net inflow into node $i$.

To enable numerical simulation and optimization, \eqref{eq:node-ode-scalar} is discretized with a fixed time step $T_s>0$ using a forward-Euler scheme:
\begin{equation}
p_i^{\,k+1}
\;=\;
p_i^{\,k}
\;+\;
T_s\,c^2\,\frac{1}{V_i}\!\left(
\sum_{e\in\mathcal E_i^{\mathrm{in}}} q_e^{\,k}
\;-\;
\sum_{e\in\mathcal E_i^{\mathrm{out}}} q_e^{\,k}
\;+\;
w_i^{\,k}
\right),
\qquad i=1,\dots,n,
\label{eq:disc-node-scalar}
\end{equation}
where $p_i^{\,k}$ is the pressure at node $i$ at step $k$, $q_e^{\,k}$ is the flow on edge $e$ at step $k$
(obtained from the edge constitutive relations), $w_i^{\,k}$ is the node injection/withdrawal at step $k$,
and $\mathcal E_i^{\mathrm{in}}$, $\mathcal E_i^{\mathrm{out}}$ are the sets of edges directed into, out of node $i$.

\subsection{Control and monitoring mechanisms}

\subsubsection{Measurement model}

In field operation a pipeline is monitored by a SCADA (Supervisory Control and Data Acquisition) system that polls pressure transmitters at selected nodes and flow meters on chosen pipe segments.  
Each scan returns a time‐stamped vector of sensor readings that the controller treats as the plant output.  
To capture this process we introduce the following measurement equation:

\begin{equation}\label{eq:stacked-meas}
\mathbf{y}_k
\;=\;
\mathbf{C}\,
\begin{bmatrix}
\mathbf{p}_k\\[2pt]
\mathbf{q}_k
\end{bmatrix}
+\mathbf{v}_k,
\qquad
\mathbf{C}=
\begin{bmatrix}
\mathbf{S}_p & \mathbf{0}\\[2pt]
\mathbf{0}   & \mathbf{S}_q
\end{bmatrix}
\end{equation}
where $\mathbf{y}_k\in\mathbb{R}^{\ell}$ is the vector of raw SCADA readings at step $k$;  
$\mathbf{p}_k$ and $\mathbf{q}_k$ are the nodal‐pressure and edge‐flow states introduced earlier;  
$\mathbf{v}_k$ represents zero‐mean measurement noise;  
$\mathbf{S}_p$ and $\mathbf{S}_q$ are binary (or scaled) selector matrices whose non-zero rows correspond to the locations of installed pressure and flow sensors.  
The block-diagonal structure of $\mathbf{C}$ makes explicit that pressures and flows are simply concatenated to ensure consistent units for subsequent state estimation and control tasks.

\subsubsection{State Estimation}

In practice the operator does not measure pressures at every node. Only a subset of pressures and a few line-flow meters are available through SCADA, and these measurements are noisy and may be delayed. Nevertheless, the supervisory controller requires an estimate of the full nodal-pressure state to enforce safety limits, run the MPC, and detect anomalies. We therefore estimate the unmeasured states with an extended Kalman filter (EKF) that blends the physics-based model with the sensor data:
\begin{equation}
\hat{\mathbf{p}}_{k+1|k+1}
\;=\;
\mathcal{E}\!\big(\hat{\mathbf{p}}_{k|k},\,\mathbf{y}_{k+1}\big)
\label{eq:estimator-def}
\end{equation}
where $\hat{\mathbf{p}}_{k|k}\in\mathbb{R}^n$ is the posterior estimate at step $k$ and $\mathbf{y}_{k+1}\in\mathbb{R}^\ell$ is the SCADA measurement vector at step $k{+}1$. The operator $\mathcal{E}(\cdot)$ denotes an EKF tailored to the discrete-time nodal-pressure model and the stacked measurement model used in this work. At each step, the EKF (i) propagates a one-step pressure prediction with the discrete-time dynamics; (ii) forms a predicted measurement by stacking selected pressures and flows (flows computed from the hydraulic/device relations); (iii) linearizes the dynamics and measurement maps at the current estimate $\hat{\mathbf{p}}_{k|k}$ via a first-order Taylor expansion, with Jacobians obtained from the same valve-conductance and compressor pressure-ratio formulas used in the model; (iv) sets the noise covariances using sensor specifications for $\mathbf{R}$ (we take $\mathbf{R}$ diagonal with entries $(0.005\,\mathrm{MPa})^2$) and tunes $\mathbf{Q}$ by innovation–covariance matching so the predicted residual variance matches the empirical one; and (v) corrects the prediction with the innovation (actual minus predicted measurements) to return $\hat{\mathbf{p}}_{k+1|k+1}$. To handle nonlinear devices, we evaluate the compressor and valve sensitivities at the operating point, clip derivatives when end-pressures are nearly equal, and freeze local slopes when an actuator is at a hard limit.

\subsubsection{Control Strategy}

Model Predictive Control (MPC) is an optimisation-based control strategy 
that, at each sampling instant, solves a finite-horizon optimal control 
problem based on a dynamic model of the system, applies the first control 
input, and repeats this process in a receding-horizon fashion.

In this paper, it is assumed that the controller predicts the evolution 
of the nodal pressures over a finite prediction horizon of length $N$ 
steps into the future. At the current time $k$, the notation $\mathbf{p}_{k+i|k}$ denotes the predicted pressure vector $i$ steps ahead,  obtained using the model and all information available at time $k$. 
For example, $\mathbf{p}_{k+1|k}$ is the one-step-ahead prediction, while $\mathbf{p}_{k+N|k}$ is the $N$-step-ahead prediction. This multi-step prediction allows the controller to anticipate future violations of 
constraints and to adjust the current control action accordingly.

At each sampling instant $k$, the supervisory controller determines the \emph{reference actuator commands} 
$\boldsymbol{\alpha}^{\mathrm{ref}}_{k} \in \mathbb{R}^{n_u}$, which specify the target settings for all controllable devices in the network 
(e.g., compressor pressure ratios, valve openings). These references are computed by solving a finite-horizon optimization problem:

\begin{equation}
\min_{\{\boldsymbol{\alpha}_{k+i}\}_{i=0}^{N-1}}
\;\;
\sum_{i=0}^{N-1}
\left\|
\mathbf{W}_p\big(\mathbf{p}_{k+i+1|k} - \mathbf{p}^{\star}_{k+i+1}\big)
\right\|_2^2
\;+\;
\sum_{i=0}^{N-1}
\left\|
\mathbf{W}_\alpha\,\Delta\boldsymbol{\alpha}_{k+i}
\right\|_2^2,
\label{eq:mpc-obj}
\end{equation}
subject to
\begin{align}
\mathbf{p}_{k+i+1|k} &=
\hat{\mathbf{p}}_{k+i|k}
+ \mathbf{A}_{k+i}\big(\mathbf{p}_{k+i|k} - \hat{\mathbf{p}}_{k+i|k}\big)
+ \mathbf{G}_{k+i}\,\boldsymbol{\alpha}_{k+i}
+ \mathbf{d}_{k+i},
\quad i=0,\dots,N-1,
\label{eq:mpc-dyn} \\[3pt]
\mathbf{p}_{\min} &\le \mathbf{p}_{k+i|k} \le \mathbf{p}_{\max},
\quad i=0,\dots,N,
\label{eq:mpc-pbounds} \\[3pt]
\boldsymbol{\alpha}_{\min} &\le \boldsymbol{\alpha}_{k+i} \le \boldsymbol{\alpha}_{\max},
\quad i=0,\dots,N-1,
\label{eq:mpc-abounds} \\[3pt]
\|\Delta\boldsymbol{\alpha}_{k+i}\|_\infty &\le r_{\max},
\quad i=0,\dots,N-1,
\label{eq:mpc-ramps}
\end{align}
where $\Delta\boldsymbol{\alpha}_{k+i} = \boldsymbol{\alpha}_{k+i} - \boldsymbol{\alpha}_{k+i-1}$.

The cost function in \eqref{eq:mpc-obj} consists of two terms. The first penalizes deviations of predicted pressures $\mathbf{p}_{k+i+1|k}$ from the desired nominal profile $\mathbf{p}^{\star}_{k+i+1}$, with $\mathbf{W}_p$ specifying the relative importance of each pressure component. The second term penalizes actuator changes $\Delta\boldsymbol{\alpha}_{k+i}$, with $\mathbf{W}_\alpha$ controlling the smoothness of compressor ratio and valve opening adjustments. 

Constraint~\eqref{eq:mpc-dyn} comes from the discretised and linearised nodal pressure dynamics.  
It ensures that the predicted pressures over the MPC horizon evolve according to the approximated system model,  
linking current pressures, control inputs, and known disturbances.  
This constraint is needed so that the optimisation respects the pipeline’s physical behaviour while planning control actions. Constraint \eqref{eq:mpc-dyn} enforces consistency between the predicted pressures and the underlying system dynamics over the prediction horizon. It is obtained by linearizing the discrete-time nodal pressure update equation \eqref{eq:disc-node-scalar} around the latest state estimate and nominal control input. The matrices $\mathbf{A}_{k+i}$ and $\mathbf{G}_{k+i}$ represent the Jacobians of the pressure dynamics with respect to pressure and actuator input, respectively, and $\mathbf{d}_{k+i}$ collects known terms such as forecasted withdrawals. By imposing this constraint, the optimizer ensures that all predicted pressure trajectories are physically feasible under the local linear model, enabling real-time optimization while preserving model fidelity. Constraint \eqref{eq:mpc-pbounds} imposes lower and upper bounds $\mathbf{p}_{\min}$ and $\mathbf{p}_{\max}$ on nodal pressures to ensure safe operating conditions across the network. Constraint \eqref{eq:mpc-abounds} enforces physical operating limits on the actuators, with $\boldsymbol{\alpha}_{\min}$ and $\boldsymbol{\alpha}_{\max}$ defining allowable compressor ratios and valve openings. Constraint \eqref{eq:mpc-ramps} limits the maximum absolute change in any actuator between consecutive time steps, where $r_{\max}$ specifies the allowable ramp rate, ensuring smooth actuator transitions and reducing mechanical wear.

\subsection{Bi-Level Attack-Control Formulation}
\label{sec:bilevel}

We formalize the cyber--physical interaction between an adversary and the
supervisory controller as a bi-level program. The \emph{upper level} (attacker)
designs small additive signals on \emph{sensors only} (false-data injection,
FDI) to degrade service pressure at demand nodes while remaining stealthy under
the SCADA bad-data detector (BDD). The \emph{lower level} (controller) reacts
optimally by solving the MPC problem already defined in
\eqref{eq:mpc-obj}--\eqref{eq:mpc-ramps}, using the discrete-time nodal-pressure
model \eqref{eq:disc-node}, the stacked measurement model \eqref{eq:stacked-meas},
and the EKF update \eqref{eq:estimator-def}.

\begin{equation}\label{eq:bilevel}
\begin{aligned}
\underset{\{\mathbf{e}^y_k\}_{k=0}^{h}}{\text{maximize}}\;
& -\sum_{k=0}^{h}\sum_{e\in\mathcal{F}} q_{e,k} \\[6pt]
\text{subject to}\;
& \bigl\|\mathbf{S}_p\bigl(\mathbf{p}_k-\hat{\mathbf{p}}_{k|k}\bigr)+\mathbf{e}^y_k\bigr\|_2
  \le \tau_S, \qquad k = 0,\dots,h, \\[8pt]
& (\mathbf{p},\mathbf{q},\boldsymbol{\alpha})\;\in\;
  \arg\min_{\boldsymbol{\alpha}}
  \bigl\{\text{MPC problem \eqref{eq:mpc-obj}--\eqref{eq:mpc-ramps}}\bigr\}.
\end{aligned}
\end{equation}

Here, the decision variables of the upper level are the additive false-data-injection vectors on the pressure sensors, $\{\mathbf{e}^y_k\}_{k=0}^{h}$. The objective in \eqref{eq:bilevel} maximises the negative of the cumulative edge flows $q_{e,k}$ over the selected flow set $\mathcal{F}$, which is equivalent to minimising the total throughput delivered during the attack horizon $k=0,\dots,h$. The stealth constraint ensures that the attack remains undetected: $\hat{\mathbf{p}}_{k|k}$ is the EKF posterior from \eqref{eq:estimator-def}, $\mathbf{S}_p$ selects the pressure channels monitored by the BDD, and the innovation residual $\mathbf{S}_p(\mathbf{p}_k-\hat{\mathbf{p}}_{k|k})+\mathbf{e}^y_k$ must have Euclidean norm below $\tau_S$ to remain within the detector’s acceptance region. The lower level is the MPC problem from \eqref{eq:mpc-obj}--\eqref{eq:mpc-ramps}, solved at each $k$ over its prediction horizon $i=0,\dots,N-1$, producing the control sequence $\boldsymbol{\alpha}$ and the resulting state and flow trajectories $(\mathbf{p},\mathbf{q})$.

\subsection{Solving the bilevel attack and control problem}

The developed bilevel optimization problem is solved by replacing the lower level MPC with its Karush–Kuhn–Tucker (KKT) optimality conditions and thus obtaining a single level mixed integer quadratic program that can be handled by standard solvers. The MPC in \eqref{eq:mpc-obj} to \eqref{eq:mpc-ramps} is a convex quadratic program because the cost is quadratic and the linearised dynamics, pressure limits, actuator bounds, and ramp limits are affine. Stacking the horizon variables as
\begin{equation}
\mathbf{z} = \big[\;\{\mathbf{p}_{k+i|k}\}_{i=1}^{N}\;;\;\{\boldsymbol{\alpha}_{k+i|k}\}_{i=0}^{N-1}\;\big]
\end{equation}
the lower level can be written compactly as
\begin{equation}
\min_{\mathbf{z}}\;\tfrac12\,\mathbf{z}^{\!\top}\mathbf{H}\,\mathbf{z}+\mathbf{h}^{\!\top}\mathbf{z}
\quad\text{subject to}\quad
\mathbf{G}\mathbf{z}\le \mathbf{g},\;\;
\mathbf{E}\mathbf{z}=\mathbf{e},
\end{equation}
with \(\mathbf{H}\) and the matrices \((\mathbf{G},\mathbf{g},\mathbf{E},\mathbf{e})\) assembled directly from \eqref{eq:mpc-dyn} to \eqref{eq:mpc-ramps} at time \(k\).

For a convex quadratic program the KKT conditions are necessary and sufficient. Introducing multipliers \(\boldsymbol{\lambda}\ge 0\) for the inequalities and \(\boldsymbol{\nu}\) for the equalities, the KKT system is
\begin{equation}
\label{eq:kkt-lower}
\begin{aligned}
&\text{stationarity} && \mathbf{H}\mathbf{z}+\mathbf{h}+\mathbf{G}^{\!\top}\boldsymbol{\lambda}+\mathbf{E}^{\!\top}\boldsymbol{\nu}=0,\\
&\text{primal feasibility} && \mathbf{G}\mathbf{z}\le \mathbf{g},\quad \mathbf{E}\mathbf{z}=\mathbf{e},\\
&\text{dual feasibility} && \boldsymbol{\lambda}\ge 0,\\
&\text{complementarity} && \boldsymbol{\lambda}\odot\big(\mathbf{g}-\mathbf{G}\mathbf{z}\big)=\mathbf{0}.
\end{aligned}
\end{equation}
The complementarity relations are linearised with a big \(M\) formulation by introducing binaries \(\mathbf{s}\in\{0,1\}^{m_I}\) for the \(m_I\) inequality rows,
\begin{equation}
\label{eq:bigM}
0\le \boldsymbol{\lambda} \le M\,\mathbf{s},\qquad
0\le \mathbf{g}-\mathbf{G}\mathbf{z} \le M\,(\mathbf{1}-\mathbf{s}),
\end{equation}
which yields mixed integer linear inequalities coupled with the stationarity equation.

Substituting \eqref{eq:kkt-lower} and \eqref{eq:bigM} into the upper level replaces the follower’s \(\arg\min\) by its optimality conditions. The stealth requirement \(\big\|\mathbf{S}_p(\mathbf{p}_k-\hat{\mathbf{p}}_{k|k})+\mathbf{e}^y_k\big\|_2\le\tau_S\) is retained explicitly as a second order cone. The resulting single level model is a mixed integer quadratic program with a second order cone constraint in the attack variables together with the primal–dual MPC variables. For the 15-node test network examined in the case study, we employed CPLEX to solve the resulting mixed-integer program.

\section{Case Studies}

To illustrate the effectiveness of the proposed methodology, we conducted two case studies. The first involved a synthetic gas transmission subnetwork with 15 nodes, while the second used a 24-node network from the GasLib dataset \citep{Schmidt2017GasLib}. These testbeds capture key characteristics of real-world pipeline systems, yet remain computationally manageable for optimization and simulation analyses.

\subsection{Case Study 1}
\subsubsection{Network Configuration and Parameter Settings}

The test network has 15 nodes and 16 directed pipelines. It contains three upstream supply sources, three major demand sinks, and nine internal actuator/junction nodes (three compressors, three backbone junctions, and three controllable branch valves). Control elements comprise \(3\) compressors on the transmission trunks and \(3\) throttling valves located upstream of the demand centers. 

\begin{figure}[h!]
\centering
\begin{tikzpicture}[
  node/.style={circle, draw, minimum size=9.5mm, font=\small},
  pipe/.style={->, >=Latex, line width=0.9pt, shorten >=2pt, shorten <=2pt}
]


\node[node] (S1) at (0,8)  {S1};
\node[node] (C1) at (2.8,8){C1};
\node[node] (J1) at (5.6,8){J1};
\node[node] (V1) at (8.4,8){V1};
\node[node] (D1) at (11.2,8){D1};

\node[node] (S2) at (0,6)  {S2};
\node[node] (C2) at (2.8,6){C2};
\node[node] (J2) at (5.6,6){J2};
\node[node] (V2) at (8.4,6){V2};
\node[node] (D2) at (11.2,6){D2};

\node[node] (S3) at (0,4)  {S3};
\node[node] (C3) at (2.8,4){C3};
\node[node] (J3) at (5.6,4){J3};
\node[node] (V3) at (8.4,4){V3};
\node[node] (D3) at (11.2,4){D3};

\draw[pipe] (S1) -- (C1);        
\draw[pipe] (C1) -- (J1);        
\draw[pipe] (J1) -- (V1);        
\draw[pipe] (V1) -- (D1);        

\draw[pipe] (S2) -- (C2);        
\draw[pipe] (C2) -- (J2);        
\draw[pipe] (J2) -- (V2);        
\draw[pipe] (V2) -- (D2);        

\draw[pipe] (S3) -- (C3);        
\draw[pipe] (C3) -- (J3);        
\draw[pipe] (J3) -- (V3);        
\draw[pipe] (V3) -- (D3);        

\draw[pipe] (J1) to[out=-90, in=90, looseness=0.7] (J2);   
\draw[pipe] (J2) to[out= 90, in=-90, looseness=0.7] (J1);  
\draw[pipe] (J2) to[out=-90, in=90, looseness=0.7] (J3);   
\draw[pipe] (J3) to[out= 90, in=-90, looseness=0.7] (J2);  

\end{tikzpicture}
\caption{Topology of the test gas distribution network (15 nodes and 16 edges).}
\label{fig:houston-network-clean}
\end{figure}
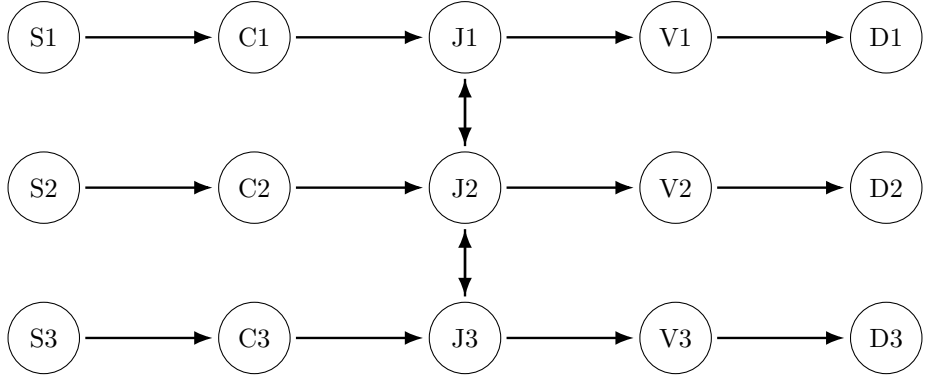

Figure~\ref{fig:houston-network-clean} shows a planar 15-node subnetwork with 16 directed pipes arranged in five left-to-right tiers: sources \((S1\text{--}S3)\), compressors \((C1\text{--}C3)\), backbone junctions \((J1\text{--}J3)\), controllable valves \((V1\text{--}V3)\), and demands \((D1\text{--}D3)\). Each row forms a trunk \(S_r \!\to\! C_r \!\to\! J_r \!\to\! V_r \!\to\! D_r\) for \(r\in\{1,2,3\}\). At the junction level, we added two sideways connections (\(J1 \leftrightarrow J2,\; J2 \leftrightarrow J3\)).  These pipes allow flow in both directions, so the different branches can share the load.

The key physical and operational parameters used in the simulation are listed in Table~\ref{tab:params}. These parameters are selected based on commonly adopted engineering practice values \citep{de2024handbook}. The system is initialized with uniform nodal pressures of 3.5~MPa and zero flows along all pipeline segments. External injections are initialized at the three supply nodes with mass-flow rates of \SI{10}{\kilogram\per\second}, \SI{12}{\kilogram\per\second}, and \SI{15}{\kilogram\per\second}, respectively. 
The SCADA system observes pressures at all demand nodes and records flows on selected transmission lines. Measurement noise is modeled as zero-mean Gaussian noise with the standard deviation given in Table~\ref{tab:params}.

\begin{table}[H]
\centering
\small
\caption{Key simulation parameters}
\label{tab:params}
\setlength{\tabcolsep}{7pt}
\renewcommand{\arraystretch}{1.1}
\begin{tabular}{@{} l c l @{}}
\toprule
\multicolumn{3}{@{}l}{\textbf{Physics and network}}\\
\midrule
Isothermal sound speed & $c$ & \SI{380}{\meter\per\second} \\
Friction factor (uniform) & $f_e$ & \num{0.012} \\
Pipe diameter (uniform) & $D_e$ & \SI{0.50}{\meter} \\
Pipe length & $L_e$ & 10 to \SI{30}{\kilo\meter} \\
Initial pressure (all nodes) & $p_0$ & \SI{3.5}{\mega\pascal} \\
\addlinespace[2pt]
\multicolumn{3}{@{}l}{\textbf{Limits}}\\
\midrule
Pressure bounds & $p_{\min},\,p_{\max}$ & \SI{2.0}{\mega\pascal}, \SI{5.0}{\mega\pascal} \\
Control bounds & $\boldsymbol{\alpha}_{\min},\,\boldsymbol{\alpha}_{\max}$ 
& compressor ratio $\in [1.0, 1.5]$; valve opening $\in [0,1]$ \\
Ramp limit (per step) & $r_{\max}$ & \num{0.05} \\
\addlinespace[2pt]
\multicolumn{3}{@{}l}{\textbf{Sensing and detection}}\\
\midrule
Pressure sensors (count) & $\ell_p$ & 6 \\
Measurement noise (pressure) & $\mathbf{R}$ & $\sigma^2\mathbf{I}$, $\sigma=\SI{0.01}{\mega\pascal}$ \\
Process noise (pressure) & $\mathbf{Q}$ & $10^{-5}\mathbf{I}$ \\
BDD residual threshold & $\tau_S$ & \SI{0.075}{\mega\pascal} \\
\addlinespace[2pt]
\multicolumn{3}{@{}l}{\textbf{Exogenous profiles and attack}}\\
\midrule
External injections/withdrawals & $\mathbf{w}_k$ & piecewise constant, $\sim\SI{10}{\kilo\gram\per\second}$ \\
Attack horizon & $h$ & 32 steps (8 h) \\
\bottomrule
\end{tabular}
\end{table}

\subsubsection{Results}

\begin{figure}[H]
  \centering
  \includegraphics[width=0.8\linewidth]{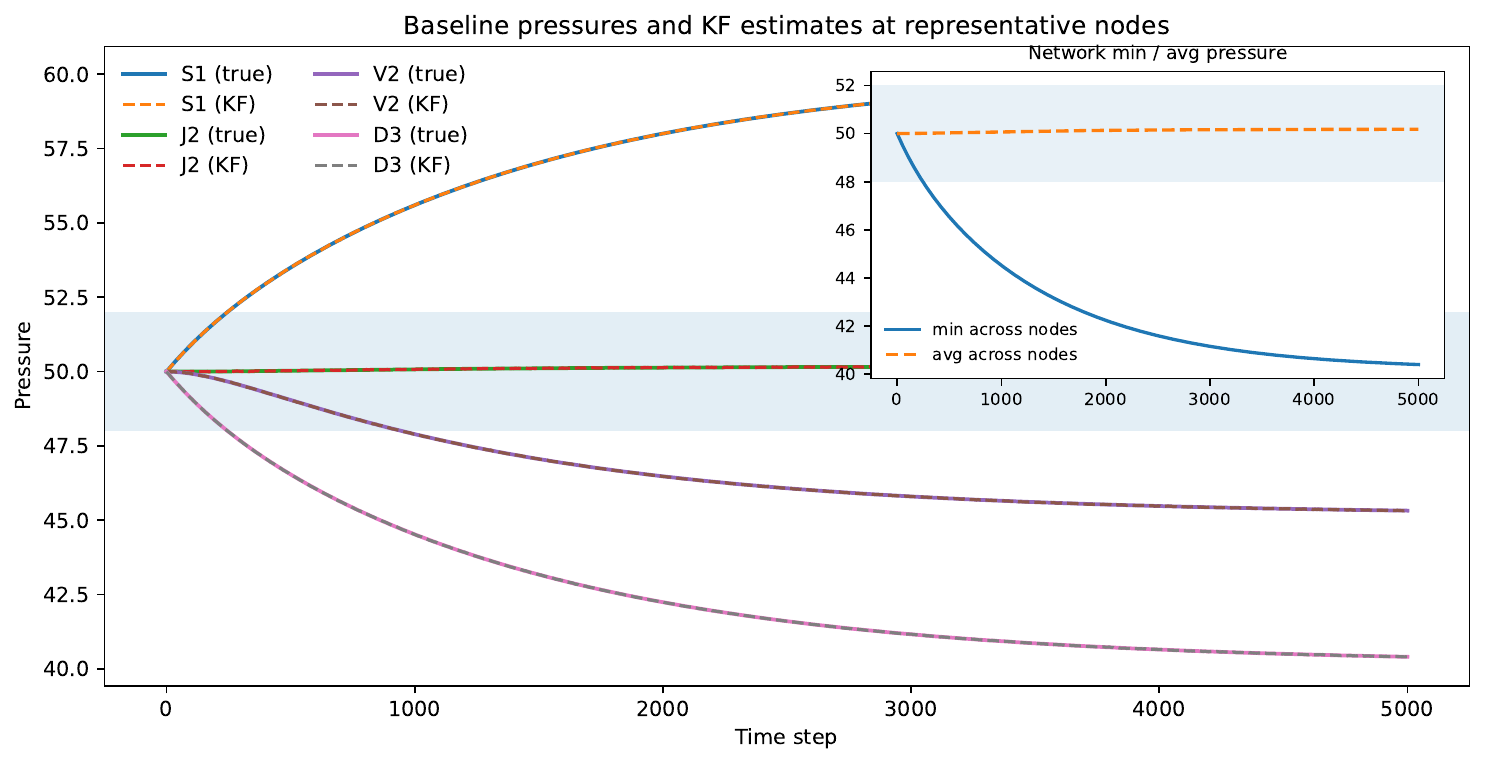}
  \caption{Pressure distribution in the test network}
  \label{fig:baseline-pressures}
\end{figure}

We evaluate the proposed estimation and control and attack framework, which includes the discrete-time network dynamics, the SCADA measurement model, the EKF update, the MPC controller, and the bi-level interaction. Figure~\ref{fig:baseline-pressures} shows pressures at four representative nodes under the baseline case, which is the normal operating condition of the network without disturbances or adversarial actions. Solid curves are the true pressures and dashed curves are the Kalman–filter (KF) estimates. The light band marks the nominal operating range. The small panel inside the figure reports the minimum and the average pressure across all nodes. At the upstream source node S1 the pressure rises gradually because sustained injections build pressure near the source and the effect propagates through the network. At the intermediate and downstream nodes J2, V2, and D3 the pressure declines as withdrawals reduce local line pack and the decrease diffuses along the pipes toward a new steady level. The close overlap of solid and dashed curves indicates that the KF tracks the state accurately at the baseline noise level. This figure is physically consistent with gas-flow behavior and shows that the estimator is reliable for our operating conditions. 
These trajectories serve as the baseline for later scenarios, where deviations from them quantify the impact on service and on estimation performance.

\begin{figure}[H]
  \centering
  \includegraphics[width=0.8\linewidth]{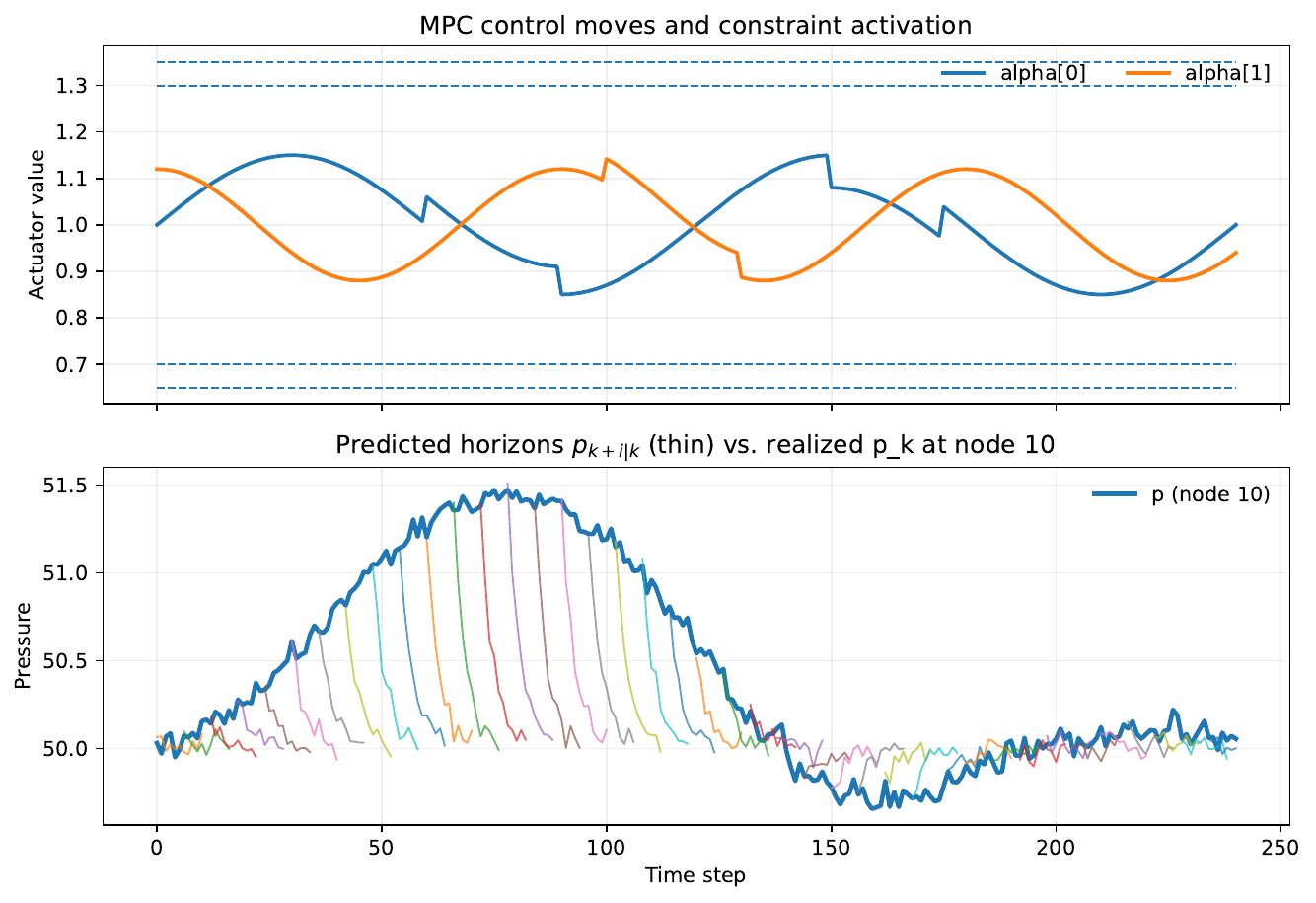}
  \caption{MPC actions and pressure predictions}
  \label{fig:mpc-ctrl-pred}
\end{figure}

Figure~\ref{fig:mpc-ctrl-pred} documents how the single MPC coordinates two actuated devices while predicting and regulating pressure at a representative location. The upper panel plots the two control inputs computed at every sampling instant: the compressor setpoint at C2 (node 4) and the valve opening at V2 (node 10). Both trajectories remain within the prescribed bounds $(\boldsymbol{\alpha}_{\min},\boldsymbol{\alpha}_{\max})$ and satisfy the ramp limit $r_{\max}$. Short flat segments appear when a bound becomes active. Functionally, C2 raises midline pressure upstream of the demand corridor, whereas V2 throttles the branch toward D2 to shape the distribution. Their coordinated motion also redistributes flow through the lateral ties between J1, J2, and J3. 

The lower panel focuses on node 10 (V2) and compares, at every time $k$, the $N$-step-ahead pressure predictions $\{p_{k+i|k}\}_{i=1}^{N}$ (thin fans) with the realized pressure $p_k$ (solid curve). Because predictions are recomputed after each SCADA scan via the estimator, successive fans re-center around the latest state and tighten as constraints become active. The realized pressure stays inside the admissible band $[p_{\min},p_{\max}]$, with only small transients attributable to process noise and model mismatch.

\begin{figure}[H]
\centering
\includegraphics[width=0.8\linewidth]{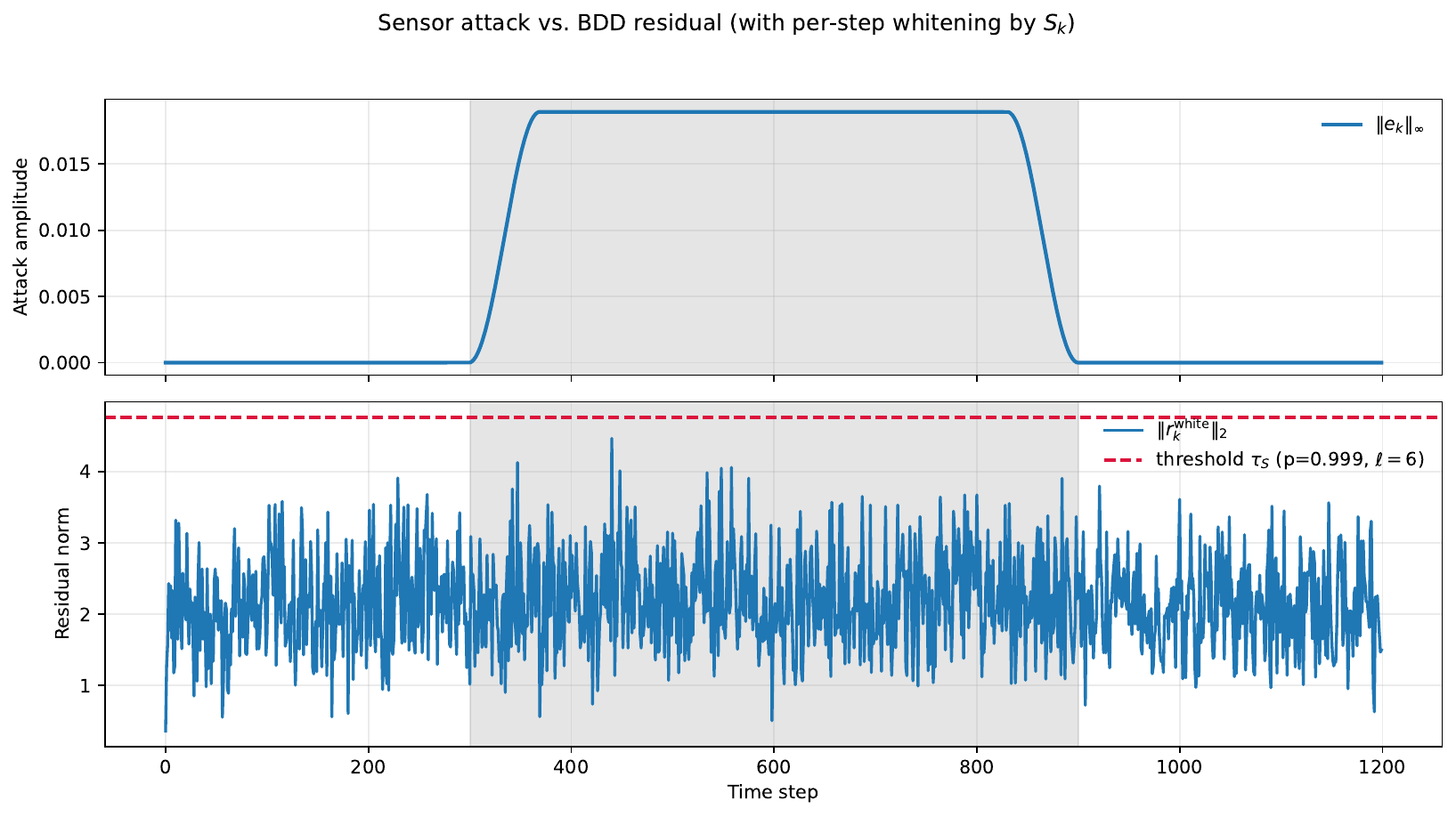}
\caption{Covert sensor attack vs. BDD residual}
\label{fig:stealth-bdd}
\end{figure}

Figure~\ref{fig:stealth-bdd} evaluates the stealthiness of the proposed bi-level attack strategy. The residuals are whitened at each time step so that their statistical properties are normalized, and the resulting test statistic is compared against a $\chi^2$-based detection threshold at a high confidence level ($p=0.999$). The lower panel shows that, during the entire shaded attack window, the residual norm consistently remains below the detection threshold, indicating that the attack is not flagged by the bad-data detector. In contrast, the upper panel illustrates that the pressure sensors are subject to a deliberate perturbation, introduced with a smooth ramp-up and ramp-down profile. This means that the attack successfully manipulates sensor readings to influence system behavior, while at the same time staying hidden within the detector’s acceptance region. Such results confirm that the proposed attack formulation satisfies the stealth requirement, which achieves covert manipulation without triggering standard anomaly detection mechanisms.

\begin{figure}[H]
  \centering
  \includegraphics[width=0.8\textwidth]{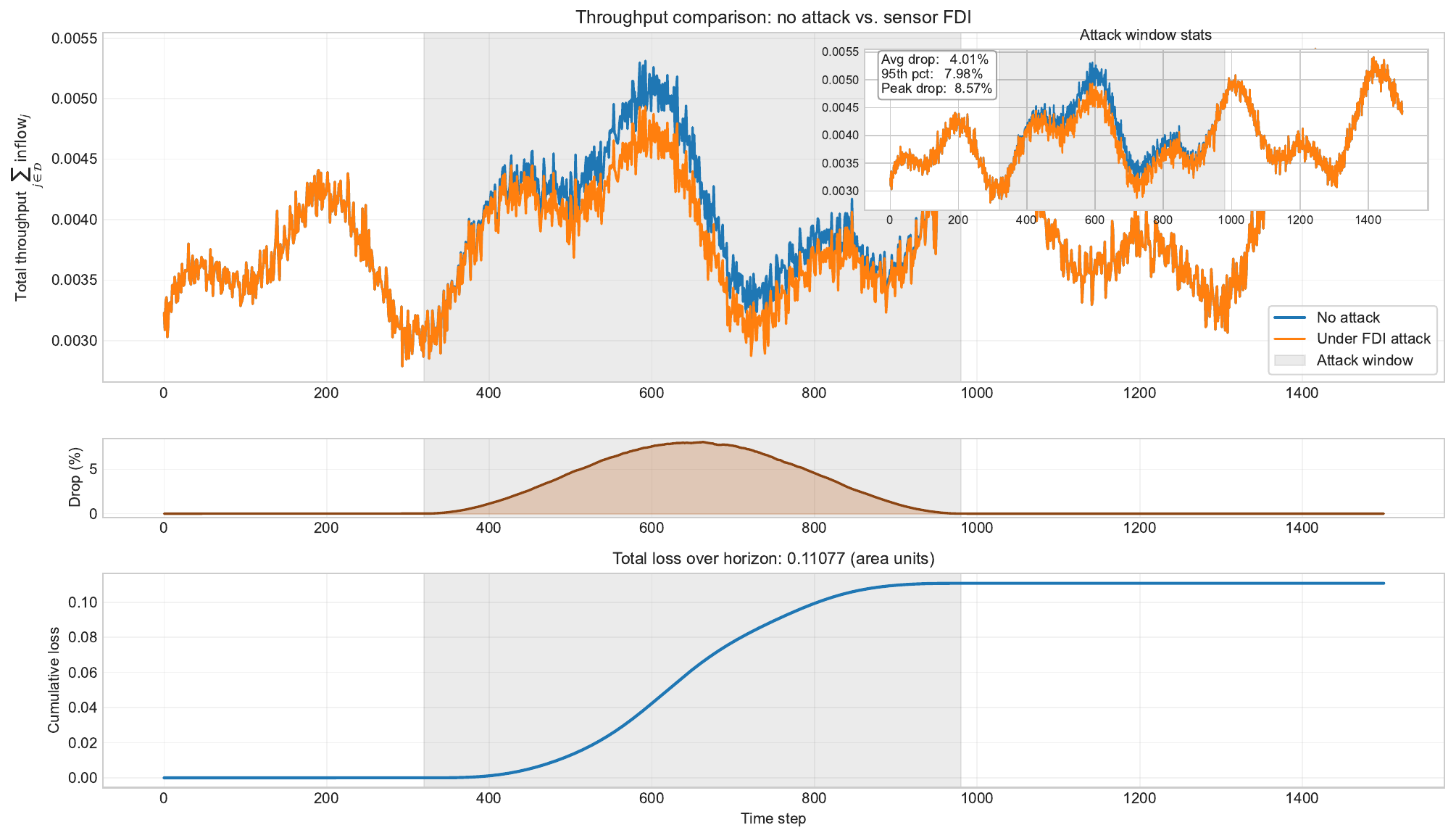}
  \caption{Network delivery comparison (nominal vs. sensor data attack)}
  \label{fig:throughput_fdi}
\end{figure}

\noindent\textbf{Figure~\ref{fig:throughput_fdi}} illustrates how a sensor data attack affects the overall volume of flow delivered by the system, expressed here as throughput. In the top panel, the blue and orange curves initially coincide, showing that under normal operation the attacked system and the baseline system deliver nearly the same output. Once the attack begins, within the shaded interval, the curves start to diverge. Although the deviation is small and not immediately obvious, the inset confirms that the average reduction in delivered flow is about four percent, with most losses below eight percent and a maximum below nine percent. This means that the attack does not create a dramatic change that would be visible to operators at a glance, but it still produces a persistent reduction in output. The middle panel summarizes this effect by plotting the smoothed percentage loss at each instant. The loss follows the same raised-cosine shape as the injected disturbance, rising gradually, reaching a peak within the attack window, and then falling back as the disturbance ends. The close alignment between the loss curve and the attack profile confirms that the degradation in service is directly caused by the manipulated sensor data. The bottom panel shows the cumulative impact of this small but sustained loss. Each short-term reduction, though modest on its own, accumulates over time to produce a noticeable deficit in total service. By the end of the simulation the area under the loss curve translates into a significant cumulative reduction. Together, these three views demonstrate that the attack produces subtle but systematic performance degradation. The effect is difficult to detect in real time because instantaneous deviations are small, yet the overall loss becomes material once the attack persists long enough.

\autoref{tab:throughput-summary} summarizes throughput under the FDI attack. Compared with the baseline, the average delivery drops by about 4\% with a peak reduction of nearly 9\%.  
The cumulative loss indicates a sustained impact over the attack window, highlighting that even stealthy attacks can cause measurable degradation in service.

\begin{table}[H]
\centering
\caption{Throughput summary}
\label{tab:throughput-summary}
\begin{tabular}{l S[table-format=1.6] S[table-format=1.6] S[table-format=2.2] S[table-format=2.2] S[table-format=2.2] S[table-format=1.6]}
\toprule
{Case} & {Baseline} & {Attacked} & {Mean} & {Peak} & {Median} & {Cumulative} \\
 & {mean [units/s]} & {mean [units/s]} & {drop [\%]} & {drop [\%]} & {drop [\%]} & {loss [area]} \\
\midrule
FDI & 0.004044 & 0.003876 & 4.15 & 8.57 & 4.02 & 0.110773 \\
\bottomrule
\end{tabular}
\end{table}

\subsection{Case Study 2}
\label{sec:case24}

\subsubsection{Network Configuration and Parameter Settings}
\begin{figure}[H]
\centering
\includegraphics[width=\linewidth]{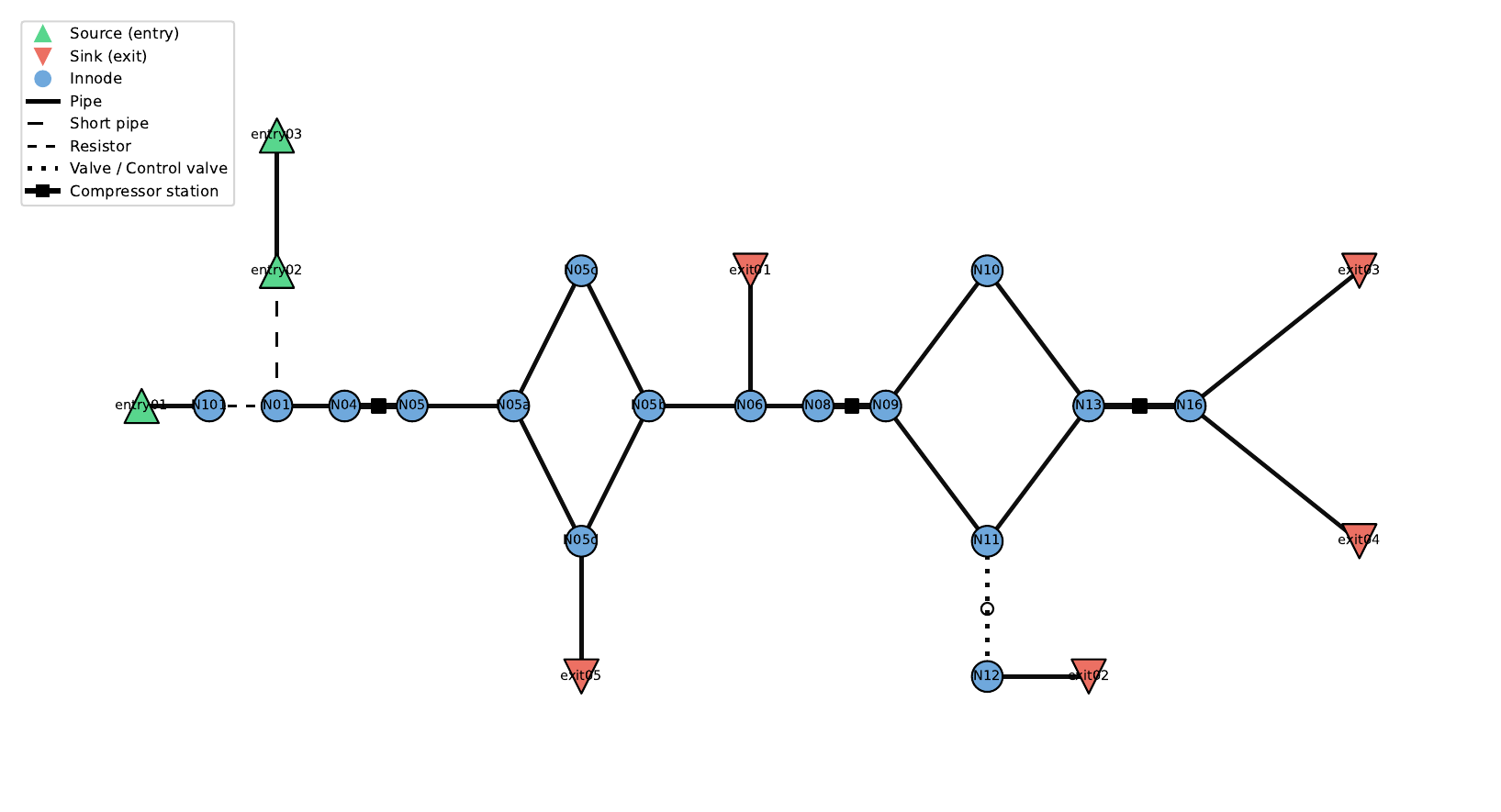}
\caption{24-node network used in the case study. Blue circles: junctions; solid lines: pipes; dashed styles: short-pipe/resistor segments; dotted line with marker: valve/control valve; black squares: compressor stations; green triangles: sources (entries); red inverted triangles: sinks (exits).}
\label{fig:net24}
\end{figure}

In this case study, we use the GasLib 24-node dataset \citep{Schmidt2017GasLib}, which provides realistic topologies and device classes derived from European pipeline data. The network consists of 24 nodes and 34 interconnecting pipes, including three supply (entry) nodes, five demand (exit) nodes, and 16 junctions. Four edges are actively controllable: three compressor stations and one control valve. For monitoring, we assume a representative SCADA subset of sensors, including pressure transmitters at selected nodes and flow meters on selected lines, as shown in Figure \ref{fig:net24}. 

Table~\ref{tab:params24} summarizes the parameters used in the GasLib-24 case study. Where available, parameter ranges (e.g., device classes, pipe diameters/lengths) follow the public GasLib data. The remaining values (e.g., noise levels) use standard engineering settings for simulation and are reported explicitly below. The physical network is modeled with an isothermal sound speed of $c=\SI{350}{\meter\per\second}$, a friction factor between $0.010$ and $0.012$, pipe diameters ranging from \SI{0.50}{\meter} to \SI{2.10}{\meter}, and pipe lengths between \SI{10}{\meter} and \SI{100}{\kilo\meter}. All nodes start at an initial pressure of $p_0=\SI{5.0}{\mega\pascal}$. 
Operational limits require pressures to stay within $[p_{\min},p_{\max}]=[\SI{3.0}{\mega\pascal},\,\SI{7.0}{\mega\pascal}]$, compressor ratios within $[1.0,\,1.60]$, valve openings within $[0,\,1]$, and actuator changes to respect a per-step ramp limit of $r_{\max}=0.10$. 
The sensing and detection setup includes $\ell_p=12$ pressure sensors, measurement noise $\mathbf{R}=\sigma^2\mathbf{I}$ with $\sigma=\SI{0.005}{\mega\pascal}$, process noise $\mathbf{Q}=(\SI{0.02}{\mega\pascal})^2\mathbf{I}$, and a bad-data detection threshold $\tau_S=\SI{0.005}{\mega\pascal}$. External injections and withdrawals $\mathbf{w}_k$ are modeled as piecewise constant profiles between \SI{5}{} and \SI{15}{\kilo\gram\per\second}.

\begin{table}[H]
\centering
\small
\caption{Key simulation parameters for the GasLib-24 case study}
\label{tab:params24}
\setlength{\tabcolsep}{7pt}
\renewcommand{\arraystretch}{1.1}
\begin{tabular}{@{} l c l @{}}
\toprule
\multicolumn{3}{@{}l}{\textbf{Physics and network}}\\
\midrule
Isothermal sound speed & $c$ & 350m$s^{-1}$ \\
Friction factor (typical) & $f_e$ & 0.010-0.012 \\
Pipe diameter (range) & $D_e$ & 0.50m - 2.10m \\
Pipe length (range) & $L_e$ & 10m - 100km  \\
Initial pressure (all nodes) & $p_0$ & 5.0 MPa \\
\addlinespace[2pt]
\multicolumn{3}{@{}l}{\textbf{Limits}}\\
\midrule
Pressure bounds & $p_{\min},\,p_{\max}$ & 3.0 MPa, 7.0 MPa \\
Control bounds & $\boldsymbol{\alpha}_{\min},\,\boldsymbol{\alpha}_{\max}$ 
& compressor ratio $\in [1.0,\,1.60]$; valve opening $\in [0,\,1]$ \\
Ramp limit (per step) & $r_{\max}$ & \num{0.10} \\
\addlinespace[2pt]
\multicolumn{3}{@{}l}{\textbf{Sensing and detection}}\\
\midrule
Pressure sensors (count) & $\ell_p$ & 12 \\
Measurement noise (pressure) & $\mathbf{R}$ & $\sigma^2\mathbf{I}$,\; $\sigma=\SI{0.005}{\mega\pascal}$ \\
Process noise (pressure) & $\mathbf{Q}$ & (0.02MPa$)^2\,\mathbf{I}$ \\
BDD residual threshold & $\tau_S$ & 0.005 MPa \\
\addlinespace[2pt]
\multicolumn{3}{@{}l}{\textbf{Exogenous profiles and attack}}\\
\midrule
External injections/withdrawals & $\mathbf{w}_k$ & piecewise constant,\; \(\sim\)5 - 15kgs$^{-1}$ \\
Attack horizon & $h$ & 20 steps \\
\bottomrule
\end{tabular}
\end{table}

\subsection{Results}

\begin{figure}[H]
  \centering
  \includegraphics[width=\linewidth]{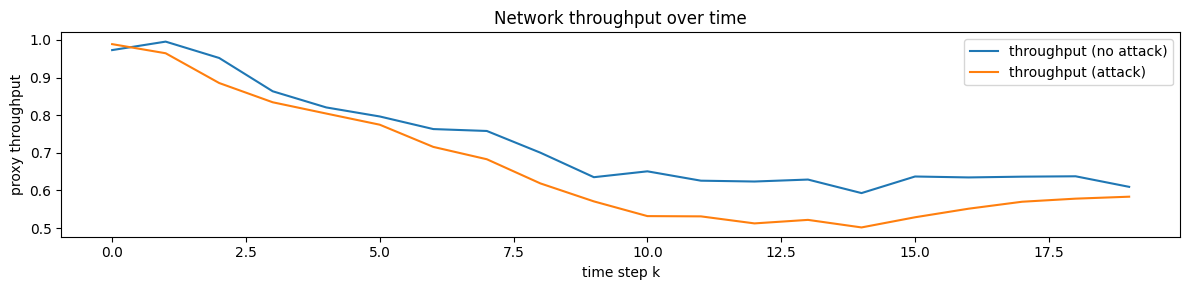}
  \caption{Network throughput over time with and without attack.}
  \label{fig:throughput}
\end{figure}

\autoref{fig:throughput} shows that the attack reduces delivery. The no-attack curve stays above the attacked curve for most of the horizon. The gap grows after a few steps and then narrows slightly near the end. This indicates the stealth FDI biases the estimator enough to steer MPC to less favorable operating points while keeping constraints satisfied.

\begin{figure}[H]
\centering
\includegraphics[width=\linewidth]{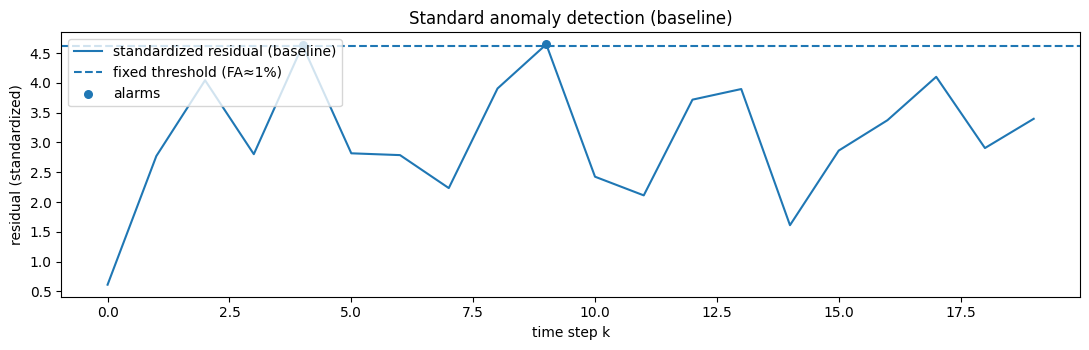}
\caption{Standard detection under a stealthy attack: standardized residual vs. fixed threshold}
\label{fig:rmse-kf-robust}
\end{figure}

Figure~\ref{fig:rmse-kf-robust} evaluates a standard anomaly-detection baseline under the same stealthy FDI sequence generated by our bilevel attack design. The detector is the conventional residual-threshold test layered on a standard Kalman filter: at each time step we compute the measurement-prediction mismatch, normalize it by its predicted uncertainty, and declare an alarm only when this standardized residual exceeds a fixed threshold. The threshold is calibrated on no-attack data to achieve a target false-alarm rate of about 1\% and then kept constant for the entire run. All settings includign nodal model, sensor placement, MPC inputs, noise statistics, and initialization are identical to those used in previous example. The figure indicates that most standardized residuals are below the set threshold. There are only occasional instances where this threshold is exceeded. As a result, the baseline detector does not effectively identify the bilevel-designed attack. This confirms the stealth property of our attack relative to a widely used detection strategy.

\begin{figure}[H]
\centering
\includegraphics[width=\linewidth]{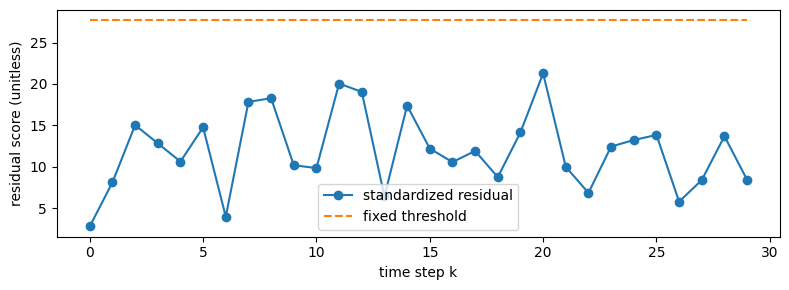}
\caption{Residual score and fixed threshold under DoS attack}
\label{fig:bdd-vs-gate}
\end{figure}

Figure~\ref{fig:bdd-vs-gate} shows our framework under a denial-of-service (DoS) measurement-dropping attack. We use the same network, sensors, and control settings as in the FDI study, and generate the DoS sequence within the same bilevel optimization framework. In this case, the DoS attack operates by randomly dropping half of the sensor measurements at each time step. The figure reports a residual score over time (a unitless measure of the mismatch between measurements and model predictions) together with a fixed threshold calculated from no-attack data. Under the optimized DoS policy, the residual score stays below the threshold at almost all steps, so a standard residual-threshold detector would not raise alarms. This shows that our bilevel design produces stealthy and effective attacks beyond additive FDI, which extends to availability-type disruptions such as DoS.

\subsection{Computational performance and scalability}

We evaluated the KKT-based MIQP on two networks (15-node and GasLib-24) under identical solver settings (CPLEX~22.1.1, MIP gap $=1\%$, time limit $=5400$~s, 2 threads) on Google Colab. For each configuration we ran 20 trials with different noise seeds and report median and interquartile ranges. On the 15-node case the median per-solve time is about 42~minutes; on GasLib-24 the median time is about 60 minutes with a final optimality gap under 1\%. This longer runtime mainly stems from the MIQP’s branch-and-bound over many binary decisions (from the KKT/complementarity reformulation) being solved with limited threads.

\section{Conclusions}\label{conclusions}

This paper presented a physics informed modeling and optimization framework to analyze cyber induced impacts on gas pipeline operations. The network was represented on a graph with nodal pressure dynamics and edge flow relations of Weymouth type, augmented with control aware elements such as valves and compressors. A SCADA measurement model and an extended Kalman filter were used to reconstruct unmeasured pressures and flows, which enabled model predictive control to compute actuator commands under pressure limits, actuator bounds, and ramp constraints. Adversarial manipulation was formulated as a bilevel problem in which an attacker perturbs sensor readings while remaining below a bad data detection threshold, and the controller responds by solving an optimal control problem. The attacker controller interaction was reformulated via KKT conditions into a single mixed integer quadratic program. Two case studies were conducted. One involved a network with 15 nodes. The other involved a network with 24 nodes. The case studies showed that sensor level attacks can stay statistically undetected yet cause persistent throughput reduction with small instantaneous deviations. The case studies employ simplifying assumptions, such as isothermal flow, uniform friction factor, and constant diameter, to enhance clarity. However, these assumptions may restrict direct application to real-world scenarios. Future work could address this limitation by incorporating real-world data.

Future work will focus on three areas: (1) The physical modeling will be improved by adding more realistic features such as temperature changes, elevation effects, gas composition variations, and more accurate equations of state; (2) The control and attack strategies will be expanded so that control will be made more robust using advanced model predictive control methods that can handle uncertainty and errors in state estimation. The attack model will cover more complex threats, including coordinated attacks on sensors and actuators, replay attacks, denial-of-service events, and protocol manipulation, even under limited attacker knowledge; (3) To ensure practical use, future work will focus on making the method faster and scalable using techniques like decomposition, warm-starting, and parallel computing. The framework will also be tested on large-scale, realistic pipeline systems using actual SCADA data and operator-in-the-loop studies to support real-world risk assessment and guide better system design decisions. Moreover, the proposed bilevel framework assumes the attacker’s model matches the real system. In practice, the attacker may have an imperfect model, which usually reduces attack impact and can even make detection easier. To address this, future work will (i) relax the perfect-knowledge assumption by introducing model uncertainty into the upper level and testing attacks with misspecified dynamics, and (ii) explore defender strategies that exploit mismatch, such as parameter variation or adaptive thresholds.

\section{Acknowledgments}
This work was supported by the US Department of Transportation (USDOT) Tier-1 University Transportation Center (UTC) Transportation Cybersecurity Center for Advanced Research and Education (CYBER-CARE). The USDOT UTC Award Number: 69A3552348332.

\bibliographystyle{unsrtnat}
\bibliography{references}

\end{document}